\documentclass{elsart}

\usepackage{natbib}

\usepackage{graphicx}
\usepackage{amssymb}

\begin{document}

\begin{frontmatter}



\title{Neutron Stars and Gamma Ray Bursts with LOFAR}%\thanksref{footnote1}}
%\thanks[footnote1]{This template can be used for all publications in Advances in Space Research.}


\author{Joeri van Leeuwen}%\corauthref{cor}\thanksref{footnote2}}
\address{Stichting ASTRON, PO Box 2, 7990 AA Dwingeloo, The Netherlands}
%\corauth[cor]{Corresponding author}
%\thanks[footnote2]{Additional information regarding the corresponding author}
\ead{leeuwen@astron.nl}
%url can be given like this
%\ead[url]{http://authors.elsevier.com/locate/latex}

\author{The LOFAR Transients Key Science Project}%%Second Author and Third Author\thanksref{footnote3}}
\address{http://www.transientskp.org/}
%% \thanks[footnote3]{Additional information about the second and third authors}
%% \ead{more@email.addresses}
%% 
%% \author{More Authors\thanksref{footnote4}}
%% \address{Address of the co-authors}
%% \thanks[footnote4]{Additional information about the co-authors}
%% \ead{more@email.addresses}

\begin{abstract}
% Text of abstract

LOFAR, the Low Frequency Array, is an innovative new radio telescope
currently under construction in the Netherlands. With its continuous
monitoring of the radio sky we expect LOFAR will detect many new
transient events, including GRB afterglows and pulsating/single-burst
neutron stars.  We here describe all-sky surveys ranging from a time
resolution of microseconds to a cadence span of years.

\end{abstract}


\end{frontmatter}

\parindent=0.5 cm

% main text
\section{Introduction}

New possibilities for neutron-star (NS) and gamma-ray burst (GRB)
research are emerging from radio interferometry using large numbers of
low-cost receivers, such as in the currently operational ATA
\citep{bow07b} and the planned SKA \citep{kram04}. We here investigate
the prospects of finding neutron stars -- be it classical radio pulsars,
AXPs or RRATs -- and GRBs with LOFAR (for details see
\citealt{ls09}). Besides these the {\em Transients Key Science
  Project} aims to study all variable and transient LOFAR sources, including jet sources such as AGN or supernovae as well as
accreting white dwarfs, neutron stars and stellar-mass black
holes. Sources will also include exo- or solar-system planets, flare
stars, brown dwarfs, and active binaries; as well as short radio
bursts, generally unexplored parameter space, ETI and the unknown
\citep[for an overview, see][]{fws+08}.

\section{LOFAR - The Low Frequency Array}

With the first stations operational and the first pulsars detected
LOFAR is on track to officially start operation early 2010. Using two different types of dipoles, LOFAR can observe in a low and a
high band that range from 30-80\,MHz and 110-240\,MHz
respectively. The sensitivity using the high-band antenna (HBA) is
several times that of the low-band antenna (LBA) so we will here
briefly outline the HBA characteristics.

Sets of 4x4 dipoles form an antenna 'tile'.  Tiles are grouped together in
stations that increase in collecting area with their distance from the
array center. The
innermost 6 split stations are are packed tightly in a
'superstation'. Spread over the 2-km core there are 12 more split HBA
stations. Next are 18 Dutch
'remote', while the $\sim$8 large international stations are spread over
Europe. Signals from these stations are next sent to the
central processor supercomputer for correlation, addition and/or
different types of beam forming. 

\section{Neutron Stars}

In \citet{ls09} we have investigated the number and type of radio pulsars
that will be discovered with LOFAR. We consider different search
strategies \citep[such as ``coherent'' versus ``incoherent'',
  cf.~][]{bac99} for the Galaxy, for globular clusters and for other
galaxies. We show that a 25-day all-sky Galactic survey can find
approximately 900 new pulsars, probing the local pulsar population to
a very deep luminosity limit. For targets of smaller angular size such
as globular clusters and galaxies more LOFAR stations can be combined
coherently, to make use of the full sensitivity. Searches of nearby
northern-sky globular clusters can find new low luminosity millisecond
pulsars (eg.\ several millisecond pulsars in a 10-hour observation of
M15). Giant pulses from Crab-like extragalactic pulsars can be
detected out to over a Mpc.

This survey will produce a complete local census of radio-emitting
neutron stars, such as radio pulsars and AXPs and with the long
pointings possible it is particularly sensitive to transient-type
neutron stars like RRATs \citep{mll+06} and intermittent pulsars
\citep{klo+06} -- for an overview see \citet{hsl+09}. This census
provides insight into neutron birth rates and properties, hence
elucidating core collapse energetics and asymmetry, the velocities and
spatial distribution of pulsars, and the physics of neutron stars in
general.

\begin{figure*}[]
   \includegraphics[width=\textwidth]{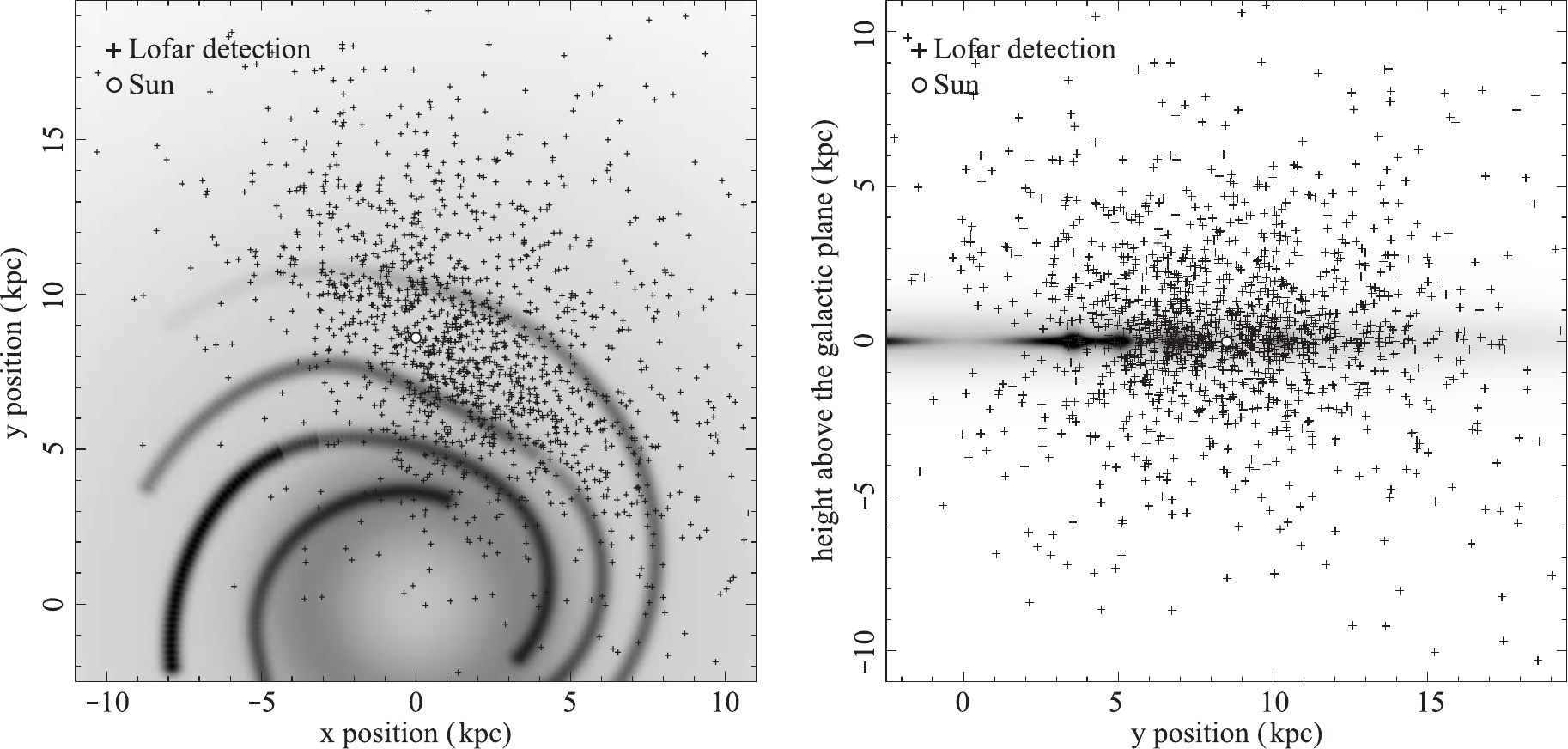}
	  \caption{The 1100 simulated pulsars observable
  with the reference Galactic LOFAR survey, for 1-hour pointings. The scattering
  interstellar matter (ISM) is shown in gray. {\bf Left)} projected on the
  Galactic plane. {\bf Right)} projected on the plane through the Galactic
  centre and sun, perpendicular to the disk.  The Galactic centre is
  at (0,0,0). LOFAR probes the local
  population very well, while the broadening from multi-path
  scattering on the ISM limits what volume
  a low-frequency survey like LOFAR probes. 
  \protect\citep[ISM modeled after][]{tc93}. \vspace{10mm}
  \label{img:x-y-z} 
}
\end{figure*}

\section{Gamma Ray Bursts}

Prompt emission and (orphan) afterglows from GRBs can be picked up in
one of several Transients KSP surveys: in the high time resolution
(pulsar, exoplanet) dedicated all-sky survey; in the piggyback survey
that will search all LOFAR observations to look for variable and
transient sources, by comparing with previous images of that region of
sky; in targeted deep high-resolution observations \citep[cf.\ GRB
  030329][]{hkr+08} triggered by other
facilities such as orbiting X-ray/$\gamma$-ray observatories, groundbased
optical telescopes; or in the dedicated Radio Sky Monitor, rapid
regular scans of a large fraction of the entire northern sky that we
will describe in some more detail below.

\begin{figure}[]
  \includegraphics[width=\textwidth]{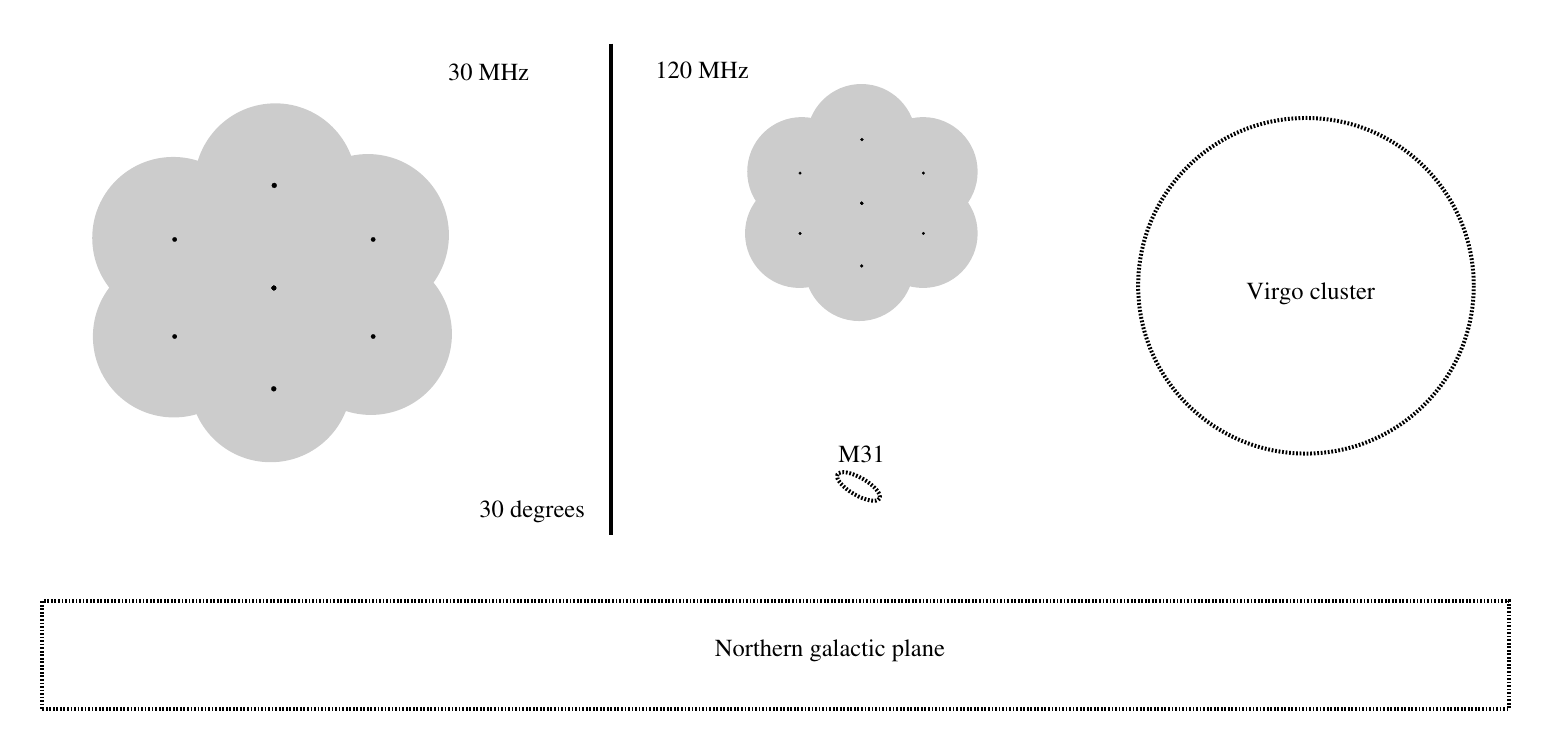}
  \caption{ Comparison of approximate Radio Sky Monitor (RSM) fields
    of view and the angular size on the sky of M31, the Virgo cluster
    and the northern galactic plane. Figure from
    \protect\citet{fws+08}. \vspace{10mm}
    \label{img:rsm}
}
\end{figure}

The RSM could operate in several modes. Rapid All-Sky monitoring mode:
fast shallow ($\sim$1mJy) hemispherical surveys could be performed on
short timescales ($\sim$1min pointings, daily) in order to survey for
rapid transients. Zenith monitoring mode: staring at the zenith
optimises the sensitivity and beam stability of the telescope, whilst
providing a sizeable and repeatedly monitored part of the sky.
Galactic plane monitoring: most of the northern galactic plane is
visible from LOFAR and could be monitored for galactic transients.

\newpage 
Every second calibrated images are produced of each beam. These are
accumulated over logarithmic intervals to integrated maps, and
analysed in real-time to check for flux changes of known objects or
for new transients. High-confidence events will initiate LOFAR follow
up observations as well as trigger alerts directly to partner
observatories, and to the broader community.

% \appendix

\end{document}